\documentclass[plb,aps,twocolumn,showpacs]{revtex4-1} 
\usepackage{graphicx}
\usepackage{epstopdf}

\begin{document}

\oddsidemargin = -12pt

\title{Algebraic models for shell-like quarteting of nucleons}

\author{
J. Cseh \\
Institute for Nuclear Research, Hungarian Academy of Sciences,
                Debrecen, Pf. 51, Hungary-4001} 
 
\date{\today}

\begin{abstract}
Algebraic models are proposed for the description of the shell-like quarteting
of the nucleons both on the phenomenologic and on the semimicroscopic levels.
In the previous one the quartet is considered as a structureless object, while
in the latter one its constituents are treated explicitly. The excitation spectrum 
is generated by the SU(3) formalism in both cases. An application to the $^{20}$Ne
nucleus is presented.
\end{abstract}





\maketitle

\oddsidemargin = -12pt


Quarteting is an important phenomenon in several branches of physics
\cite{wu,bose}.
In nuclear physics it appears in a straightforward way due to the fact that the
exclusion principle allows two (spin ${1 \over 2}$) protons and neutrons to occupy
a single-particle state, and the
short-range  attractive  nucleon-nucleon forces prefer this arrangement.
Therefore, it has long been known,
and recently a conjecture was put forward on the importance of quarteting 
also in nuclei away from the line of stability
\cite{nature}.

A well-known signature of quarteting is that
the separation energy of a nucleon in an 
even-even $N=Z$ nucleus is much larger than that 
of an $\alpha$-particle. The fact that the nuclear
mass of $4n$ nuclei is approximately a  linear function of $n$,
while the masses of $4n+x$ nuclei are quadratic
function of $x$, was already the motivation for Wigner's
supermultiplet theory
\cite{wigner}. 
(Much work has been done on the binding energies and quarteting
later on, too,  see e.g. 
\cite{pviq}.)

A nuclear quartet model was formulated in
\cite{stretch},
(based on the stretched scheme),
and then it was generalized in several steps.
In 
\cite{arima}
quartet excitations   were considered from
one major shell to the other, 
and the corresponding energies were determined
from mass relationships. In this generalized
interpretation a quartet is not related to a specific
angular momentum coupling scheme: it is 
made of 2 protons 
and 2 neutrons, occupying a fourfold degenerate
single particle state ($l,m$ orbit in $L-S$ coupling,
or $j,m$ and $j,-m$ orbits in $j-j$ coupling).
The internal binding  of a quartet is strong,
while the quartet-quartet interaction is relatively 
weak.  Arima and Gillet took into account 
\cite{roton}
also pairs of nucleons,  as further building blocks, 
extending the description to even-even nuclei of different $Z$ and $N$. 

In
\cite{satpat}
intrashell quartet excitations have been introduced
in addition to the intershell excitations of
\cite{arima}.
This concept leads to a quartet shell model, 
i.e. one assumes the existence of a 
self-consistent quartet potential well,
and its states are used to describe the quartet
states in $4n$ nuclei. The 
$0s$, $0p$, $1s-0d$, ...
oscillator shells of the nucleon-shell model
are replaced by 
$0s$, $0p$, $1s-0d$, ...
quartet shells, having 1, 3, 6, ... single quartet states, 
respectively.
The corresponding energies were determined empirically, too.

A further extension was presented in
\cite{harvey} 
by incorporating any number of particle-hole
excitations (in the language of the nucleon-shell-model),
contrary to the quartet-shell-model of 
\cite{arima,satpat}
which had only 0, 4, 8 ... excitation quanta
(in terms of nucleon-shell-model).
This considerable extension of the quartet model space appeared
due to  the conceptual generalization of a quartet. Harvey defined 
\cite{harvey}
it as  2 protons and 2 neutrons having a quartet-symmetry:
permutational symmetry of [4],
and spin-isospin symmetry of [1,1,1,1].

Interacting boson type quartet models were invented 
\cite{dukel, iachjack}
for the description of quarteting in heavy nuclei.
In 
\cite{dukel} the basic building block quartets
are treated as   $l=0$ ($s$) and $l=2$ ($d$) bosons, 
and the model has a U(6) group structure,
like the interacting boson model of the quadrupole collectivity
\cite{ibm}.
This model describes 
a spectrum of positive parity states.
In
\cite{iachjack}
the alpha-like correlation is treated in terms of bosons of
nucleon-pairs, but in addition to the 
$s$ and $d$ bosons  another set of basic
building blocks of $l=0$ ($s^*$) boson and $l=1$ ($p$) boson
is included, therefore, negative parity states are also involved.
These phenomenological models have the efficiency and elegance
of the algebraic methods in generating the spectrum. E.g. they have
dynamical symmetries as limiting cases, which provide us with exact
solutions for the eigenvalue problem.

In 
\cite{gamb}
a BCS-like study was carried out for bosons of the
proton-neutron interacting boson model
\cite{ibm}
and it was concluded that the superfluid condensate is more
of a quartet type, rather than separate superfluid phases of
proton and neutron pairs.
Recent investigations
\cite{sand,samb}
show that the isovector pairing in self-conjugate nuclei are
of quartet type and can be well described by a quartet condensation model.

Another condensate, namely the alpha-particle condensate attracts much
attention these years
\cite{acondens}.
For the first sight it is very different from the quartet condensate.
The latter one was shown to be important in the ground state, while
the Bose-Einstein condensate (BEC) of alpha particles was invented for
the description of the gas-like dilute structure near the alpha-threshold.
However, later calculations revealed that the the THSR wavefunction,
which is applied in the alpha-condensate studies have a very large
overlap with the (resonating group method) wavefunction of ground
state
\cite{horicond}, 
indicating that the overlap with the quartet condensate is
considerable, too. The non-localized nature of clustering in the BEC
also shows in this direction. The exact relation of these two condensates still
remains to be understood. In the present work we do not investigate the
condensates, rather we concentrate on the ``individual'' quartet-excitations,
in the sense of
\cite{arima,satpat,harvey}.

We propose algebraic quartet models based on the
concepts of shell-model-like quarteting of
\cite{arima,satpat,harvey}.
Our main purpose is the description of  the excitation spectrum.
We propose two models: the simpler one is called phenomenologic
algebraic quartet model (PAQM), which has
the building blocks very similar to that of the quartet-shell model of
\cite{satpat}, i.e. the composite nature of the quartet do not
appear explicitly.   The second one is the semimicroscopic algebraic
quartet model (SAQM), based on the quartet concept of 
\cite{harvey}, in which each of the four nucleons of the quartet
is treated.
The novel feature in comparison with the works 
\cite{arima, satpat,harvey}
is that
an algebraic framework is formulated for the description of
the detailed spectrum, like in the group theoretical approach of
the works
\cite{dukel,iachjack,ibm}.
On the other hand, the new models are different from the
interacting boson type  models of
\cite{dukel,iachjack},
because of the nature of their building blocks, and shell-like
structure of the model spaces.

We apply Elliott's SU(3) scheme 
\cite{elliott,harveyell}
for  generating the spectrum  both in the phenomenological and
in the semimicroscopical descriptions. In the former case structureless
quartets are supposed to occupy the single-particle levels of the
harmonic oscillator shells,  while in the latter model nucleons do so.
Therefore, the phenomenological model space has only a spatial
part, while the semimicroscopical one contains a space and a spin-isospin 
components. In fact, this latter model space 
is a  truncation of that of the $L-S$ coupled no-core shell model
\cite{nocore},
based on the spin-isospin formalism.
The physical operators are expressed in terms of the group generators,
thus  algebraic techniques can be applied in calculating the matrix elements.
\\

\noindent
{\bf The phenomenologic algebraic quartet model.}
In this approach an excitation quantum 
$(\hbar \omega)_q$ between the major shells  
is expected to be 
approximately 4 times
that of the nucleon shell model:
$(\hbar \omega)_q \approx 4 (\hbar \omega)$.
All the shells have positive parity, due to their quartet nature.
If a single quartet state is occupied, then no other
particle can be put there, therefore, the permutational 
symmetry of the quartets has to be that of a single-columned
Young diagram: [1,1,...]. 

The building blocks of the description are the nine operators,
$
{\hat A_{ \alpha \beta} } = {1 \over 2} ({\hat a^{\dag}_{ \alpha}} 
{\hat a_{ \beta}}  + {\hat a_{ \beta}}  {\hat a^{\dag}_{ \alpha }} ) \ , \
\alpha , \beta = x,y,z,
$
$
{\hat  a_{ \alpha}} = \sum _j {\hat a_{ \alpha}}(j) \ , \
{\hat  a^{\dag}}_{ \alpha} = \sum _j {\hat a^{\dag}_{ \alpha}}(j) \ , \
j = 1,...,N; \
$ 
(here $N$ is the total number of particles), 
which are number-conserving bilinear products of the creation and
annihilation operators of oscillator quanta. They can be rewritten into
three spherical tensors: a scalar operator ${\hat n}$, which is the number
of oscillator quanta,
five components of the quadrupole momentum ${\hat Q_m}$
(acting in a single major shell),
and three components of the angular momentum ${\hat L_m}$.
The nine operators {${\hat n}, {\hat Q_m}, {\hat L_m}$}  generate the U(3) group,
the eight operators {${\hat Q_m}, {\hat L_m}$}  generate the SU(3) group, and
the three {${\hat L_m}$}  are generators of the SO(3) group.

The basis states are characterized by the representation
labels of the group--chain:
\begin{eqnarray}
 U(3) \supset SU(3) \supset SO(3) \supset SO(2)
 \nonumber \\
 \vert  [n_1,n_2,n_3] , (\lambda , \mu ) \ ,  K ,L \ \ \ \ ,\ M \ \rangle .
 \label{grch}
\end{eqnarray}
Here $ n= n_1 + n_2 +n_3 $ is the eigenvalue of the ${\hat n}$ operator.
The angular momentum content of a $(\lambda ,\mu )$ representation is as
follows 
\cite {elliott,harveyell}: 
$L= K, K + 1,...,K + max {(\lambda , \mu)}$, 
$ K = min {(\lambda , \mu )},
 min {( \lambda , \mu )}  - 2,..., 1 \ or \ 0,$
with the exception of $K_L = 0$, for which
$ L = max {(\lambda , \mu)}, max {(\lambda , \mu)} - 2,..., 1 \ or \ 0 $.
The SU(3)
content is given by the 
U(k) $\supset$ SU(3)
decomposition
\cite{unu3}, 
where $k=3,6,10,...$ for the major shell
with $1,2,3,...$ quartet excitations. The irreducible 
representation (irrep) of U(k) 
is the same as that of the permutational group in the 
major shell in question. The U(3) symmetry of the
whole nucleus is obtained as a direct product of the 
major shell U(3) irreps.
The  irreps of the 
spurious center of mass excitations can be 
determined easily, due to the fact that the c.m. excitation operator
is is fully symmetric in particle indices, and has an 
 $[1,0,0]$ U(3) irreducible tensor character
\cite {es55,bn63,he71}.
We illustrate here the construction of the model space with the
 lowest-lying  states  of the $^{20}$Ne nucleus. 
The ground state contains the filled-in
0 and 1 $\hbar \omega_q$ major shells, and 1 quartet in the 
2 $\hbar \omega_q$ major shell: $(0)^1 (1)^3 (2)^1 $. The permutational
symmetries in the three subsequent major shells are:
$[1] \otimes [1,1,1] \otimes [1] $, which give the 
$ [1,1,1,1,1] $ symmetry of the five-quartet-state.
The corresponding U(3) symmetries are unique and simple in these 
cases, and  they result in a single U(3) irrep:
$[0,0,0] \otimes [1,1,1] \otimes [2,0,0] = [3,1,1]$.
The $ 1 \hbar \omega_q$ excitations are obtained in two different
ways: 1a: $(0)^1 (1)^2 (2)^2 $, or 1b: $(0)^1 (1)^3 (3)^1 $.
The permutational symmetries are: 
1a: $[1] \otimes [1,1] \otimes [1,1] = [1,1,1,1,1] \oplus ... $, and
1b: $[1] \otimes [1,1,1] \otimes [1] = [1,1,1,1,1] \oplus ... $.
The corresponding U(3) symmetries, and their products are:
1a: $[0,0,0] \otimes [1,1,0] \otimes [3,1,0] = 
[4,2,0] \oplus [4,1,1] \oplus [3,2,1]$;
1b: $[0,0,0] \otimes [1,1,1] \otimes [3,0,0] = [4,1,1] $.
In total the U(3) irreps are:
$[4,2,0] \oplus [4,1,1]^2 \oplus [3,2,1]$.
The spurious excitation of the centre of mass:
$[3,1,1] \otimes [1,0,0] = [4,1,1] \oplus [3,2,1]$.
Therefore, the real $ 1 \hbar \omega_q$ excitations are:
$[4,2,0] \oplus [4,1,1]$.

Table I.  shows the model space of $^{20}$Ne
for the 0-1 major shells  (both for the phenomenologic
and for the semimicroscopic approach).
Note here the small angular momentum content of the PAQM  space
(limited by the SU(3) quantum numbers).

\begin{table}
\caption{
SU(3)  quantum
numbers  of the states of $^{20}$Ne for the 0 and 1 major shells  in the
phenomenologic and semimicroscopic algebraic quartet model. 
The superscripts indicate multiplicity.
}
\begin{center}
\begin{tabular}{|c|c|c|}
\hline
\hline
\multicolumn{1}{|c|}{model}
&\multicolumn{1}{|c|}{$\hbar \omega$}
& \multicolumn{1}{|c|}{SU(3)} \\
\hline
\hline
PAQM&0&(2,0)\\
\cline{2-3}
&1&(2,2),(3,0)\\
\hline
\hline
SAQM&0&(8,0),(4,2),(0,4),(2,0)\\
\cline{2-3}
&1&(8,2),(9,0),(6,3),(7,1)$^2$,(4,4),(5,2)$^4$,(2,5),(6,0)\\
&&(3,3)$^4$,(1,4)$^2$,(4,1)$^3$,(2,2)$^4$,(0,3)$^2$,(3,0)$^3$,(1,1)$^2$\\
\hline
\hline
\end{tabular}
\end{center}
\end{table}

The operators of
physical quantities are obtained in this description in terms of the
generators of the U(3) group.
In particular 
the Hamiltonian can be expanded in terms of the generators
of the U(3) group, coupled to spherical scalars.
The general solution of the eigenvalue problem than involves two steps:
{\it i)} calculation of matrix elements of the Hamiltonian between the
basis states, and {\it ii)} numerical diagonalization of the energy
matrix.
In the special case of the dynamical symmetry, 
i.e. when the Hamiltonian is expressed in terms of
the invariant operators of the group-chain
(\ref{grch}),
an analytical solution is
available.

\begin{figure}
\includegraphics[height=8.1cm,angle=0.]{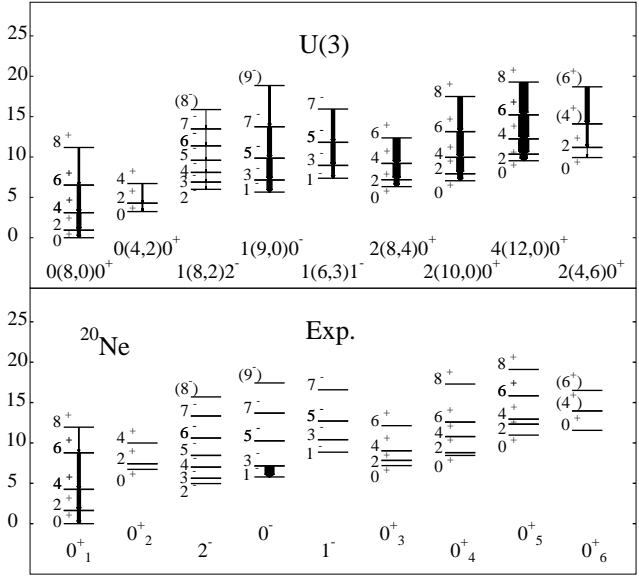}
\caption{ 
The spectrum of the semimicroscopic algebraic quartet model in
comparison with the experimental data of the  $^{20}$Ne nucleus.
The experimental bands are labeled by the $K^{\pi}$, and the model states
by the $n (\lambda, \mu)K^{\pi}$ quantum numbers. 
The spin-parity in parenthesis indicates uncertain band-assignment.
The width of the arrow between the states is proportional to the
strength of the $E2$ transition.
\label{fig:spectrum}}
\end{figure}

The electromagnetic transition operators are obtained as Hermitian
combinations of group generators with appropriate tensorial character.
The lowest--order transition operators are:

\begin{equation}
{\hat T^{(E0)}} =
e^{(0)} {\hat n} , \ 
{\hat T^{(E2)}_m}  = 
e^{(2)} {\hat Q^{(2)}_{m}} , \  
{\hat T^{(M1)}_m}  = 
m^{(1)} {\hat L^{(1)}_m} .
\end{equation}
These operators are diagonal in the SU(3) basis, i.e. they indicate no transitions
between states of different SU(3) irreps. This is a typical situation in the
algebraic models in the dynamical symmetry limit. Transitions between e.g.
different major shells can be obtained either i) by applying symmetry-breaking
interactions, which mix the SU(3) basis states, or ii) by constructing more
complex operators.  

The PAQM  states of Table I. correspond to very highly excited states
(due to the  large excitation quantum of the quartets:
$(\hbar \omega)_q \approx 4 (\hbar \omega)$).
Therefore, it is hard to find a well-established correspondence
between the experimental and model states.
In case of the semimicroscopic description, on the other hand,
it is much more straightforward.

The relation of the PAQM to the previous models is as follows.
Its model space  is identical with that of
\cite{satpat}.
The main difference between  the approach of
\cite{arima, satpat}
and the present one is that in  
\cite{arima, satpat}
the interaction matrix elements are obtained empirically from
the binding energies, while here we construct all the physical operators
algebraically. This enables us to calculate the complete spectrum in
an easy way.

\noindent
{\bf The semimicroscopic algebraic quartet model.}
On the semimicroscopic level we
take into account  the composition of the quartets explicitly.
They are considered
\cite{harvey}
as 2 protons and 2 neutrons having 
permutational symmetry of [4],
and spin-isospin symmetry of [1,1,1,1]. 
Therefore, in this case the nucleon shell model space 
is applied, and it is truncated according to these symmetries.
Subsequent major shells have opposite parities.

The building blocks of this description are, again, 
the creation and annihilation operators of oscillator quanta 
(of the nucleon shell model).

The spectrum is determined by the U(3) spatial and
the U$^{ST}$(4) spin-isospin irreps.
In this case, however, the particles are nucleons,
not structureless quartets, as in the phenomenologic
model. 
Therefore, the groups describe the
symmetries of the many-nucleon-systems.
Their relevant irreps are obtained in the following way.
In each major shell those U(k) (or permutational)
symmetries has to be taken into account
which result in the required quartet symmetry
$[4]^{N_q}$ when calculating their outer product with
those of the other major shells. (Equivalently,
those U$^{ST}$(4) irreps are relevant, which
result in the $[1,1,1,1]^{N_q}$ quartet symmetry
in the direct products.) These  U(k) 
(or U$^{ST}$(4)) symmetries determine the relevant
U(3) representations, and their direct products
define the model space.
The center of mass excitations can be removed in the same
way, like in the previous case.

The model space is much richer than that of the phenomenological model,
as shown by Table I.
Especially remarkable is the angular momentum content of the model space;
already in the lowest-lying major shell L=8 appears.

The physical operators are expressed in this case, too, in terms of the group-generators.
Due to the restriction to the quartet symmetry only the scalar
U$^{ST}$(4) part of the spin-isospin sector gives contribution to the
Hamiltonian and to the transition operators. Therefore, 
the formulae of the 
phenomenologic quartet model are valid here, too,
but the oscillator quanta in this case refer to
those of the nucleon shell model.

{\it As an  application} we show here the result of the semimicroscopic model for
the $^{20}$Ne nucleus. The U(3) dynamical symmetry approach is used;
the interactions are written in terms of the invariant operators of group-chain (1),
therefore,  an analytical solution is available. 

The experimental data are taken from 
\cite{till},
but for the band-assignment of the highly-excited alpha-cluster states
also the conclusions of
\cite{rich}
are taken into account.
The lower part of Figure 1. shows the states with definite band-assignment.
All the bands with $K^{\pi}$ values of 
\cite{till}
are included, except the one of the very uncertain (and somewhat contradictory)
$0^+_7$ band. 
In case of the $7^-$  state of the $0^-$ band, and 
the $6^+$ and $8^+$ states of the $0^+_6$ band, 
which have more than one experimental candidates,
the average energies are indicated.
(In
\cite{till}
there are only three states, which are not included here, for
not having corresponding states in the model spectrum:
a $6^+$ state in the  $0^+_2$ band,  
a $9^-$ state in the  $1^-$ band,  and  
a $8^+$ state in the  $0^+_6$ band.
Each of them have uncertain band-assignment.)

We have tried a few phenomenological interactions,
expressed in terms of the invariant operators of the 
U(3)$\supset$SU(3)$\supset$SO(3)
algebra-chain. 
Each of them contained a harmonic oscillator term
(linear invariant of the U(3)), with a strenght obtained from
the systematics
\cite{moli}
$\hbar \omega$ = 13.19 MeV, and a rotational term with a parameter 
to fit. The remaining parts were written in terms of the second 
(${\hat C}^{(2)}_{SU3}$)
and third order 
(${\hat C}^{(3)}_{SU3}$)
invariant of the SU(3). The former one accounts for the 
quadrupole-quadrupole interaction, and the latter one distinguishes 
between the prolate and oblate shapes. The simple linear combination
$
a{\hat C}^{(2)}_{SU3} + b{\hat C}^{(3)}_{SU3}
$ 
contains two parameters.
Another two-parameter term can be written as
$
{g \over {2c}} exp(-c{\hat C}^{(2)}_{SU3} -1),
$ 
which is very similar to the  $exp (- c {\hat Q} {\hat Q} -1)$ 
term of the symplectic model Hamiltonian
\cite{symp},
accounting for a set of  many-body interactions  with well-defined relative weights.
In order not to destroy the shell structure for the case of large 
excitations, the major-shell-average of the quadratic invariant 
$
(\langle {\hat C}^{(2)}_{SU3}  \rangle )
$
can be subtracted
\cite{aver}.
We have obtained the best description
(from among the (2+1) parameter formulae)
with 

\begin{equation}
{\hat H} =
(\hbar \omega) {\hat n}  +
a{\hat C}^{(2)}_{SU3} +
({\hat C}^{(2)}_{SU3} - \langle {\hat C}^{(2)}_{SU3}  \rangle ) +
b{\hat C}^{(3)}_{SU3} +
d {1 \over {2\theta}}{\hat L}^2,
\end{equation}
where $\theta$ is the moment of inertia calculated classically for the rigid shape
determined by the U(3) quantum numbers (for a rotor with axial symmetry)
\cite{arxiv}.
Note that the third term does not introduce new fitting parameter,
but a constant coefficient of 1 MeV is quietly understood here.
The model spectrum of Figure 1 was obtained with the parameters:
$ a = -1,065$ MeV,
$b =-0,000360$ MeV
$d = 0,808$ MeV.

We note here that the experimentally identified bands are described by the
lowest-lying models bands with the appropriate spin-parity content, i.e.
the other model bands of the same character are all higher-lying.

The intraband E2 transition rates were calculated with the operator of Eq. (4).
The $B(E2)$ value is given by the formula
\cite{sun}:
\begin{eqnarray}
B(E2, I_i \rightarrow I_f) =
\ \ \ \ \ \ \ \ \ \ \ \ \ \ \ \ \ \ \ \ \ \ \ \ \ \ \ \ \ \ \ \ \ \ \ \ \ \ \ \ \ 
\nonumber \\
{{2I_f +1} \over {2I_i +1}} {\alpha}^2
\vert \langle  (\lambda , \mu) K I_i , (11)2 \vert \vert (\lambda , \mu) K I_f \rangle  \vert ^2
 C(\lambda , \mu) ,
\label{e2matrix}
\end{eqnarray}
where
$
\langle  (\lambda , \mu) K I_i , (11)2 \vert \vert (\lambda , \mu) K I_f \rangle
$
is the SU(3) $\supset $ SO(3) Wigner coefficient
\cite{prog}, 
and $\alpha$ is a parameter
fitted to the  the experimental value of the
$2^+_1 \rightarrow 0^+_1$ transition of 20.3 W.u. 
The interband transition rate is zero.
\\

The relation of the SAQM to the approach of 
\cite{harvey}
is similar to that between the PAQM and the previous models of
\cite{arima,satpat}.
In particular, the two model spaces are identical, 
but the physical operators are not. 

\noindent
{\bf To sum up:} 
In this paper we have introduced two algebraic models for
the shell-like quarteting of nucleons. The simpler one is based on the quartet-concept
of Arima et al.
\cite{arima,satpat}, 
which does not treat explicitly the degrees of freedom of the constituent nucleons.
Nevertheless, the Pauli-principle is not violated in this phenomenological description,
either: the quartets of four nucleons occupy different single-particle space-states.
The semimicrosopic model is more detailed. It is based on the definition of
quartets in terms of two protons and two neutrons of  [4] permutational symmetry
\cite{harvey}.
This model is able to take into account 0, 1, 2, 3, 4,... (nucleonic) major shell
excitations, as opposed to the ``giant'' quartet excitations of the phenomenologic
approach
\cite{arima,satpat}, 
which correspond
 to $4q$, $q=0,1,2,...$ nucleon excitation quanta.
For both description the U(3) formalism of Elliott
\cite{elliott,harveyell} is applied for the calculation of the spectrum.
The semimicroscopic model is practically a symmetry-dictated truncation of
the $L-S$ coupled no-core shell model, focusing on the spin-isospin-zero sector, and
multiple excitations. It can be considered as an effective model in the sense of
\cite{effective}:
the bands of different quadrupole shapes are described by their lowest-grade
U(3) irreps without taking into account the giant-resonance excitations, built
upon them, and the model parameters are renormalised for the subspace of the
lowest U(3) irreps. 

From the viewpoint of their group-theoretical formalism these models are similar to 
the fully algebraic interacting-boson-like quartet models of the 1980'th
\cite{dukel,iachjack}, but they are different concerning the physical nature of
the quartets.

Both of these models are easy to apply, yet the semimicroscopic approach
seems to be detailed enough to account for a considerable amount of the
experimental spectrum, as illustrated by the application to the $^{20}$Ne nucleus.
We expect that in addition to its applicability to the s-d shell nuclei
it can also be extended  to the 
mass region of A=92-100 of current experimental interest.

{ Further generalizations} are possible by
 applying symmetry-breaking interactions, which result  in e.g. nonvanishing interband transitions.
Since the shell-truncations scheme of the semimicroscopic approach is based on the
nucleonic degrees of freedom it could be exported also to non-alpha-like nuclei.

As for the {\it connection to other models}, the transparent symmetry properties
of the present approach is very helpful.
Via its obvious shell-model relation,  the connection of the quartet model to the cluster and 
collective models is also well-defined
(see e.g. 
\cite{arxiv,arxiv2}, and references therein for a recent discussion). 
In this respect the models with algebraic structure are relevant,
in particular, the microscopic cluster model applying U(3) basis
\cite{hori,hechtet},
and the semimicroscopic algebraic cluster model
\cite{sacm},
as well as the symplectic shell model
\cite{symplectic},
and the contracted symplectic model 
\cite{contr} 
of the quadrupole collectivity.

Especially promising can be the application of
the present semimicroscopic quartet model in combination with the
concept of the multichannel dynamical symmetry
\cite{musy},
when  the spectra of different cluster configurations are
obtained from the quartet spectrum by simple projections.
\\

\noindent
{\it This work was supported by the OTKA (Grant No K106035).
The technical help of Mr. G. Riczu is kindly acknowledged.}



\end{document}